\documentclass{JHEP3}
\usepackage{latexsym,graphicx,epsfig,psfrag,here}
\newcommand{\M}{{\cal M}}
\newcommand{\ol}{\overline}

\newcommand{\ve}{\varepsilon}

\newcommand{\dsum}{\displaystyle \sum} 
\newcommand{\dprod}{\displaystyle \prod} 
\newcommand{\bda}{\begin{\displaymath}\begin{array}{rl}}
\newcommand{\eda}{\end{array}\end{displaymath}}
\newcommand{\be}{\begin{equation}}
\newcommand{\ee}{\end{equation}}
\newcommand{\bdm}{\begin{displaymath}}
\newcommand{\edm}{\end{displaymath}}
\newcommand{\bea}{\begin{eqnarray}}
\newcommand{\eea}{\end{eqnarray}}

\newcommand{\nn}{\nonumber \\}

\newcommand{\wt}{\widetilde}

\newcommand{\Mvariable}{}

\newcommand{\qq}{(p_a\cdot p_b)}
\newcommand{\qqo}{p_a\cdot p_b}
\newcommand{\glan}{\begin{equation}}
\newcommand{\glaus}{\end{equation}}
\newcommand{\glanf}{\begin{eqnarray}}
\newcommand{\glausf}{\end{eqnarray}}
\newcommand{\glei}[1]{Eq. (\ref{#1})}
\newcommand{\pfeil}{\rightarrow}

\newcommand{\raum}{\;\;\;\;\;}
\newcommand{\ii}{\mbox{i}}
\newcommand{\iii}{\mbox{\tiny i}}

\newcommand{\ub}{\bar{u}}
\newcommand{\Ub}{\bar{U}}
\newcommand{\gk}{\mbox{\Large (}}
\newcommand{\gK}{\mbox{\Large )}}
\newcommand{\sgk}{\mbox{\huge (}}
\newcommand{\sgK}{\mbox{\huge )}}
\newcommand{\gke}{\mbox{\Large [}}
\newcommand{\gKe}{\mbox{\Large ]}}
\newcommand{\sgkg}{\mbox{\huge \{}}
\newcommand{\sgKg}{\mbox{\huge \}}}
\newcommand{\kk}{\langle}
\newcommand{\KK}{\rangle}
\newcommand{\gmu}{\hat{\Gamma}^\mu}

\newcommand{\gnu}{\hat{\Gamma}^\nu}

\newcommand{\glambda}{\hat{\Gamma}^\lambda}

\newcommand{\sig}{\bar{\sigma}}

\newcommand{\Tr}{\,\mbox{Tr}\,}

\newcommand{\dx}{d^4x}
\newcommand{\dy}{d^4y}
\newcommand{\dz}{d^4z}
\newcommand{\ddx}{d^dx}

\newcommand{\dpp}{d^4p_a}
\newcommand{\dPP}{d^4p_b}

\newcommand{\aaa}{_a}
\newcommand{\bbb}{_b}
\newcommand{\ppm}{p_a^\mu}
\newcommand{\PPm}{p_b^\mu}
\newcommand{\ppn}{p_a^\nu}
\newcommand{\PPn}{p_b^\nu}
\newcommand{\ppl}{p_a^\lambda}
\newcommand{\PPl}{p_b^\lambda}

\newcommand{\ddk}{\frac{d^dk}{(2\pi)^d}\,}

\newcommand{\dmu}{\partial^\mu}
\newcommand{\dmuu}{\partial_\mu}
\newcommand{\dnu}{\partial^\nu}

\newcommand{\lam}{\hat{\Lambda}}

\newcommand{\MM}{M^2-\ii\,\epsilon}
\newcommand{\MMM}{M^2-\iii\,\epsilon}

\title{The one-loop functional of chiral SU(2)
\thanks{Work supported in part by a Facultas Research Fellowship of
the Univ. of Vienna and by TMR, EC-Contract  
No. ERBFMRX-CT980169 (EURODA$\Phi$NE).}}
\author{R. Unterdorfer\\ Institut f\"ur Theoretische Physik, Universit\"at 
Wien, Boltzmanngasse 5, A-1090 Vienna, Austria\\ E-mail: unterd@thp.univie.ac.at}
\abstract{The one-loop functional for chiral SU(2) of Gasser and Leutwyler is extended to
include up to three propagators.
With this generating functional it is possible to calculate all amplitudes up to order $\phi^6$
to next-to-leading order in the low-energy expansion.
An external pion is of order $\phi$ and a vector gauge boson is of order $\phi^2$. A check of the
existing amplitudes for $\gamma\gamma \to \pi\pi$ is given.
The one-loop amplitudes and the cross sections
for four-pion production in $e^+e^-$ annihilation are calculated.}
\preprint{UWThPh-2002-03}
\begin{document}
\section{Introduction}
In the low-energy region 
it is not possible to calculate cross sections by means of perturbative QCD
because of the high value of the strong coupling constant.
One has to use an effective theory. The corresponding theory in this context is called
chiral perturbation theory (CHPT) \cite{wein,gass1,gass2,gass3}.
It incorporates all symmetries of QCD with the assumption of spontaneous breakdown of
chiral symmetry (which is suggested by phenomenological and theoretical evidence \cite{ecker1}),
but no other model dependent features. CHPT is nonperturbative in the sense that
it is not an expansion in powers of the QCD coupling constant, but an expansion in powers of
the external momenta.

For a theory with a spontaneously broken symmetry the
Goldstone theorem \cite{gold} predicts the existence of massless particles, called Goldstone bosons. In
case of  CHPT these bosons are not massless
because the chiral symmetry is not an exact symmetry, as the quarks have masses. The pions are the by far
lightest pseudoscalar mesons and therefore they are identified as
the Goldstone bosons of  broken chiral SU(2).
In CHPT the asymptotic states are not quarks, but pions (or in general pseudoscalar mesons).
A spontaneously broken symmetry corresponds to a nonlinear realization a la Nambu-Goldstone
\cite{nolin}.

In the calculation of amplitudes one has to include loops to increase the precision and
to satisfy unitarity and analyticity \cite{wein}.
CHPT is not renormalizable in the sense that each loop in the perturbation expansion
produces ultraviolet divergences, which have to be renormalized with the help of counterterms
of the appropriate order in the external momenta. The finite parts of the coupling constants
occurring in the counterterms are not determined by chiral symmetry and must be
obtained from experiment or calculated using resonance exchange or other assumptions.
Each successive loop and the associated counterterms correspond to successively higher
powers of momenta. The higher loops are expected to give only small corrections at
appropriately low energies.

Gasser and Leutwyler have calculated the generating one-loop functional
of chiral SU(2) including two propagators in the loop \cite{gass1}. With this functional one can obtain all
amplitudes with at most four external particles, where an external photon counts as two particles.
In this article the functional is extended by including a third propagator. This extension makes it
easy to analyse processes with up to six external particles. In the whole work equal masses for the two
lightest quarks are assumed.

The main purpose of calculations in the framework of pure CHPT is to analyse physics
and to investigate QCD in a region ($E << 1$ GeV) where a perturbative treatment is not possible. 
CHPT was used successfully to study the low-energy structure of, e.g., $\pi\pi$ scattering \cite{pipi} or of the pion
form factor \cite{formf}. There are also important processes with more external particles.
One example for an interesting process with six external particles is the scattering $e^+e^- \to 4\pi$,
which is analysed in Sec. \ref{ampli2}.
It is an important cross section for determining the hadronic contribution to
the anomalous magnetic moment of the muon and to the fine-structure constant at $M_Z$.

In Sec. \ref{matrix} a calculation method with external fields \cite{gass1} is introduced. The generating
functional is given in Sec. \ref{onefunk}. The amplitudes of the two channels of
$\gamma\gamma \to \pi\pi$ \cite{streuunga,streuungb,streuungc} are
presented in Sec. \ref{ampli1}. 
In Sec. \ref{ampli2} the amplitudes and cross sections of  the processes
$e^+e^- \to \pi^0\pi^0\pi^+\pi^-$ and $e^+e^- \to 2\pi^+2\pi^-$ are provided.
Sec. \ref{conclu} contains some conclusions.
\section{Matrix elements}
\label{matrix}
The Lagrangian of massless QCD does not include any terms which connect the right- and
left-handed components of the quark fields
\glan q_R=\frac{1}{2}(1+\gamma_5)q \, , \raum q_L=\frac{1}{2}(1-\gamma_5)q ~.
\glaus
Thus, the Lagrangian is invariant under chiral rotations, i.e. under two independent
U$(N_f)$ transformations (flavour rotations) of the right- and left-handed quark fields\footnote{The
flavour symmetry of massless QCD
should not be mixed up with the still existing colour gauge symmetry.}.
In case of two flavours the Noether currents associated with chiral symmetry are given by :
\glan J^{\mu 0}_I=\overline{q}_I \gamma^\mu q_I
~,\raum J^{\mu a}_I=\overline{q}_I \gamma^\mu \frac{\tau_a}{2} q_I
~,\raum I=L, R
~.\glaus
The axial and vector currents are defined through the following relations:
\glan V^{\mu a}=J^{\mu a}_R+J^{\mu a}_L \raum\mbox{vector current}
~,\glaus
\glan A^{\mu a}=J^{\mu a}_R-J^{\mu a}_L \raum\mbox{axial current}
~.\glaus
Because of Lorentz invariance and isospin symmetry there is the following
relation:
\glan <0|A^\lambda_a(x)|\pi_b(p)>=\mbox{i}\,p^\lambda \delta_{ab}F\,
\mbox{e}^{-\iii\,p\cdot x}
~.\glaus 
Since chiral symmetry is spontaneously broken
the Goldstone theorem insures that the constant $F$, which is related to the physical pion
decay constant $F_\pi=$92.4 MeV through
\glan F_\pi=F(1+{\cal O}(m_{quark}))~,
\glaus
is not zero. As a consequence the axial current, or the divergence
of the axial current, plays the role of an interpolating field for the pion.

An external photon couples to the vector current. For example, the matrix element
of  the scattering $\gamma^*  \pfeil 4\pi$ is proportional to the on-shell residue of
the corresponding singularity in the Green function
\glan
<\!0|\mbox{T}\,V^\mu_{n}(x_1)A^ \nu_{o}(x_2)A^\lambda_{p}(x_3)
A^\rho_{q}(x_4)A^\sigma_{r}(x_5)|0\!>  
\label{masterf}
~.\glaus
To get such matrix elements one can use the method of external fields \cite{gass1},
which is equivalent to the method of external currents introduced by Schwinger
\cite{schwinger}.
Extra terms are added to the QCD Lagrangian:
\glan {\cal L}_{QCD}=\overline{q}\:\ii\gamma_\mu\nabla^\mu q - 
\frac{1}{2}\,\mbox{tr}_cF_{\mu\nu} F^{\mu\nu}+V_{\mu n} v^\mu_n
+A_{\mu n} a^\mu_n-S_{n} s_n+P_{n} p_n ~.
\label{QCDlag}
\glaus
$\nabla^\mu$ is the SU(3)$_C$-covariant derivative.
The external fields
$v_\mu$, $a_\mu$, and $p$ are set to zero at the end of the calculation. The scalar field $s$
is set equal to the quark mass matrix to give masses to the light quarks, which
amounts to explicit symmetry breaking.
$S_n$ and $P_n$ are the scalar and pseudoscalar currents:
$$ S_{0}=\bar{q}q
~, $$
$$ S_{i}=\bar{q} \tau_i q\raum \mbox{for}\;\;\:i=1,2,3
~, $$
$$ P_{0}=\ii\,\bar{q}\gamma_5 q
~, $$
\glan P_{i}=\ii\,\bar{q}\gamma_5 \tau_i q\raum \mbox{for}\;\:i=1,2,3
~.\glaus
By adding those terms in \glei{QCDlag}, the global chiral symmetry  of the Lagrangian is promoted
to a local symmetry  with the appropriate gauge transformation \cite{gass1}. The external field
method allows the calculation of Green functions like (\ref{masterf}) in an easy way.
The generating functional of the Green functions is of the form:
\glan W[v, a, s, p] \doteq\, <\!0|\mbox{T}\mbox{e}^{\iii\int d^4 x\,
V_{\mu n}(x) v^\mu_n(x)+A_{\mu n}(x) a^\mu_n(x)-S_{n}(x) s_n(x)
+P_{n}(x) p_n(x)}|0\!> 
~.\glaus 
One gets the Green functions by functional differentiation with respect to external fields, for example:
\glan
\left. \frac{\delta^3 \,\ii\,W[v, a, s, p]}{\ii^3 \delta a^\lambda_p(x_1)
\delta a^\rho_q(x_2)\delta v^\mu_r(x_3)}\right|_{\stackrel
{ v=0, a=0}{\scriptscriptstyle s={\cal M}, p=0}}
=<\!0|\mbox{T}\,A_{\lambda p}(x_1)A_{\rho q}(x_2)V_{\mu r}(x_3)|0\!>  
\label{masterf2}
~.\glaus
The external fields can be interpreted as fields of external particles:
\glan v^\mu+a^\mu=-e\,Q{\cal A}^\mu
~,\glaus
\glan v^\mu-a^\mu=-e\,Q{\cal A}^\mu-\frac{e}{\sqrt{2}\sin\theta_W}(W^{\mu^+}T_++\mbox{h.c.}) 
~.\glaus
${\cal A}^\mu$ is the photon field, $W^{\mu^+}$ the W-boson field, $Q$ the
charge matrix and $T_+$ contains the relevant elements of the 
Kobayashi-Maskawa matrix. \\ \\
Because of the large value of the strong coupling constant
in the low-energy region 
it is not possible to calculate the generating
functional  by means of perturbative QCD.
So one has to use an effective Lagrangian
that includes all possible terms built out of the matrix $U(\phi)$ (which
contains the pion fields)
and the external fields that respect all existing symmetries,
namely chiral symmetry, Lorentz invariance, charge conjugation and parity.
The corresponding Lagrangians given by Gasser and Leutwyler \cite{gass1,gass2}
 are the foundation of CHPT. 
In case of SU(2) and up to order $p^4$ ($p$ stands for the external momenta)
one has:
\glan {\cal L}_2=\frac{F^2}{4}\kk D_\mu U^\dag
D^\mu U+\chi^\dag U+\chi U^\dag \KK
\label{lag2}
\, ,\glaus
\glanf
\label{lop4}
& {\cal L}_4 =& \frac{1}{4}\, l_1 \,\kk D_\mu U^\dag D^\mu U \KK^2
+\frac{1}{4} \,l_2 \,\kk D_\mu U^\dag D_\nu U \KK
\kk D^\mu U^\dag D^\nu U \KK+
\nonumber \\
& & +\frac{1}{16}\, l_3 \,\kk \chi^\dag U+\chi U^\dag \KK^2
+\frac{1}{4} \, l_4 \,\kk D_\mu U^\dag D^\mu\chi
+D_\mu U D^\mu\chi^\dag \KK+
\nonumber \\
& & +l_5 \,\kk U^\dag F_{\mu\nu}^R U F^{L\,\mu\nu}\KK
+\frac{\ii}{4}\, l_6 \,\kk F_{\mu\nu}^R\,[D^\mu U, D^\nu U^\dag]+
F_{\mu\nu}^L\,[D^\mu U^\dag, D^\nu U]\KK+
\nonumber \\
& & -\frac{1}{16} \, l_7 \,\kk \chi^\dag U-\chi U^\dag \KK^2 
+\frac{1}{4}\, h_1 \,\kk \chi\chi^\dag+\frac{1}{2}(\chi\tilde{\chi}
-\chi^\dag\tilde{\chi}^\dag)\KK+
\nonumber \\
& & -(2 h_2+\frac{l_5}{2}) \,\kk F_{\mu\nu}^R F^{\mu\nu R}
+F_{\mu\nu}^L F^{\mu\nu L}\KK
+\frac{1}{4}\, h_3 \,\kk \chi\chi^\dag
-\frac{1}{2}(\chi\tilde{\chi}-\chi^\dag\tilde{\chi}^\dag)\KK
~.\glausf
There are different possible parametrizations of $U(\phi)$, which all lead to the same on-shell
amplitudes. A convenient choice is the exponential parametrization 
\glan U(\phi)=\,\mbox{exp}\,(\,\ii\, \phi /F)\, , \raum \phi=
\tau_a\phi_a=\left( \begin{array}{cc} \pi^0 & \sqrt{2}\,\pi^+ \\ \sqrt{2}\,\pi^- & -\pi^0 \end{array} \right) \, .
\glaus
The use of the covariant derivative
\glan D_\mu U=\dmuu U-\ii\,r_\mu\,U +\ii\,U\,l_\mu\, , \raum  r_\mu=v_\mu+a_\mu\, , \raum
l_\mu=v_\mu-a_\mu
\glaus
is a consequence of adding the extra terms in \glei{QCDlag}. The effective Lagrangians (\ref{lag2},
\ref{lop4}) have the same local chiral symmetry as the extended QCD Lagrangian in (\ref{QCDlag}). 
$\chi$ and $\tilde{\chi}$ are defined as follows:
\glan \chi=2B(s+\ii\,p)
~ ,\glaus
\glan \tilde{\chi}=\tau_2\,\chi^T\tau_2
~.\glaus
$F$, $B$ and the $l_i$ are coupling constants.
The Gell-Mann-Oakes-Renner relation, which can also be derived by CHPT, yields:
\glan F^2M^2=-2\hat{m}\langle 0|\bar{u}u|0\rangle\, , \,\,\,\,\mbox{where}\,\,\,\hat{m}=(m_u+m_d)/2
~.\glaus
At a scale $\mu_{\overline{MS}}=1\mbox{ GeV }$,
$\langle 0|\bar{u}u|0\rangle$ is approximately equal
to -(230 MeV)$^3$. Thus, $\hat{m}$ is assumed to be of order $M^2$:
\glan {\cal O}(\hat{m})={\cal O}(M^2)={\cal O}(p^2)
\label{ordm}
~.\glaus
\section{Calculation of the generating functional}
\label{onefunk}
CHPT can be defined via the path integral
\glan W[v, a, s, p] \doteq {\cal N}^{-1}\int d\mu[U] \;
\mbox{e}^{\iii\!\,\int\!\ddx\,{\cal L}_{\mbox{\tiny eff}}\,(v, a, s, p)}
~.\glaus
The calculation is done in d dimensions so that dimensional regularisation can be used.

It can be shown that the loop expansion is an expansion in Planck's constant or in other words 
an expansion around the solution of the classical equation of motion  $\bar{U}=\bar{u}^2$.
The matrix $U$ is expanded with the help of a traceless Hermitian matrix $\xi$:
\glan U=u^2=\ub\,\mbox{e}^{\ii\xi}\,\ub
\label{entu}
~.\glaus
To the desired ${\cal O}(p^4)$, the loop integration must only be performed over the Lagrangian
of ${\cal O}(p^2)$.
\glan W={\cal N}^{-1}\mbox{e}^{\iii\!\,\int\!\ddx\,{\cal L}_{4}(\bar{U})}
\int\!d\mu[\xi]\;\mbox{e}^{\iii\!\,\int\!\ddx\,{\cal L}_{2}(U)}
+ \dots
\label{WFunk}
~.\glaus
$d\mu[\xi]$ is the functional measure for the matrix field $\xi$.
Planck's constant is set equal to 1. \\ \\
The Lagrangian is now expanded in $\xi$:
\glanf \int \ddx\,{\cal L}_{2}(U) & = & S_2[\bar{U}]+\frac{1}{4} F^2
\int \ddx\,\kk D^\mu (\ub^\dag\xi \ub^\dag) D_\mu (\ub\xi \ub)
-\frac{1}{2} D^\mu \bar{U}^\dag D_\mu(\ub \xi^2 \ub)+ \nonumber \\
& & -\frac{1}{2} D^\mu \bar{U} D_\mu(\ub^\dag \xi^2 \ub^\dag)
-\frac{1}{2} \xi^2(\ub \chi^\dag \ub + \ub^\dag \chi \ub^\dag)\KK
+{\cal O}(\xi^3)
~.\glausf
One has to consider the terms quadratic in $\xi$ for the one-loop calculation. 
$S_2[\bar{U}]$ is the classical action to order $p^2$.
With the definitions
\glan y_\mu \doteq \frac{1}{2}\ub^\dag D_\mu \bar{U} \ub^\dag=
-\frac{1}{2}\ub D_\mu \bar{U}^\dag \ub
~,\glaus
\glan \Gamma_\mu \doteq \frac{1}{2}[\ub^\dag,\partial_\mu \ub]
-\frac{1}{2}\ii\ub^\dag r_\mu \ub - \frac{1}{2}\ii\ub l_\mu \ub^\dag
~,\glaus
\glan
\sigma \doteq \frac{1}{2} (\ub \chi^\dag \ub + \ub^\dag \chi \ub^\dag)
~,\glaus
\glan \hat{\Gamma}^{\mu\, ab} \doteq
-\frac{1}{2}\kk[\tau^a, \tau^b]\Gamma^\mu\KK
~,\glaus
\glan \hat{\sigma}^{ab} \doteq
\frac{1}{2}\kk[\tau^a, y_\mu][\tau^b, y^\mu]\KK
+\frac{1}{4}\kk\{\tau^a, \tau^b\}\sigma\KK
~,\glaus
the generating functional of ${\cal O}(p^4)$ has the form
\glan W_4=\mbox{e}^{\iii \,(S_{2}[\bar{U}]+S_{4}[\bar{U}])}{\cal N}^{-1}\int\!d\mu[\xi]\;
\mbox{e}^{-\iii\!\,\frac{F^2}{2}
\int\!\ddx\, \xi^a(x){\cal D}^{ab}(x)\xi^b(x)}
~.\glaus
The operator ${\cal D}$ is
\glan
{\cal D}=\mbox{\bf 1}\Box+\{\gmu, \partial_\mu\} + \gmu\hat{\Gamma}_\mu
+\hat{\sigma}  
~.\glaus
The factor ${\cal N}$ is the one-loop integral with ${\cal D}$ replaced by
${\cal D}_0$ \\
\glan {\cal D}^{ab}_0=\delta^{ab}\Box
+\frac{1}{2} B\kk\{\tau^a, \tau^b\}{\cal M}\KK
~,\glaus
where ${\cal M}$ is the quark mass matrix $\hat{m}{\bf 1}$.
With use of the Gaussian formula
\glan \int d^d v\,\mbox{e}^{-\frac{1}{2}v^T A v}  
=(2\pi)^{\frac{d}{2}}(\mbox{det}\,A)^{-\frac{1}{2}}
\glaus
one gets
\glan
W_4=\mbox{e}^{\iii \,(S_{2}+S_{4})}{(\det \frac{\cal D}{{\cal D}_0})}^{-\frac{1}{2}}
~.\glaus
It is well known that one can define a generating functional $Z$ of all {\it connected}
Green functions:
\glan Z \doteq - \ii\,\mbox{ln}\,W
~,\glaus
\glan Z_4=S_{2}+S_{4}+\frac{\ii}{2}\,\mbox{ln}\det\frac{\cal D}{{\cal D}_0}
~.\glaus
With the identity ln det=Tr ln we have:
\glan Z_4=S_{2}+S_{4}+\underbrace{\frac{\ii}{2}\,\mbox{Tr}\,\mbox{ln}
\frac{\cal D}{{\cal D}_0}}_
{\displaystyle Z_{\mbox{\scriptsize one loop}}}
\label{ln}
~.\glaus
In the further calculation the operator ${\cal D}$ is split into two parts:
\glan {\cal D} = {\cal D}_0+\delta
~.\glaus
With the quantity $\bar{\sigma}$ defined via
\glan \bar{\sigma} \doteq \hat{\sigma}-2 B\hat{m}\mbox{\bf 1}
\label{sigmab}
~,\glaus
$\delta$ is of the form
\glan \delta = \{\gmu, \partial_\mu\} + \gmu\hat{\Gamma}_\mu
+\bar{\sigma}  
~.\glaus
The one-loop functional can be rewritten as
\glan Z_{\mbox{\scriptsize one loop}}=\frac{\ii}{2}\,\mbox{Tr}\,\mbox{ln}
\frac{{\cal D}_0+\delta}{{\cal D}_0}=\frac{\ii}{2}\,\mbox{Tr}\,\mbox{ln}\,
(\mbox{\bf 1}\,+\delta {\cal D}_0^{-1})=\frac{\ii}{2}\,\mbox{Tr}\,\mbox{ln}\,
(\mbox{\bf 1}\,-\delta\Delta)
\label{llll}
~.\glaus
The Feynman propagator is the inverse of $-{\cal D}_0$.
The logarithm is expanded as
\glan \mbox{ln}\,(1+x)=\sum_{n=1}^{\infty}(-1)^{n-1} \frac{x^n}{n}
~.\glaus  
Gasser and Leutwyler have calculated the functional with at most two propagators
in the loop \cite{gass1}.  In this work I extend it by including a third propagator.
The functional is then of the form
\glan Z_{\mbox{\scriptsize one loop}}=-\frac{1}{2}\,\ii\Tr(\delta\Delta)
-\frac{1}{4}\,\ii\Tr(\delta\Delta\delta\Delta)
-\frac{1}{6}\,\ii\Tr(\delta\Delta\delta\Delta\delta\Delta)+{\cal O}(\phi^8)
~.\glaus
This means the functional is of order $\phi^6$ as there are at least two external pions
on each vertex in the loop. The photon field can couple to a loop vertex without external pions
and is therefore counted as order $\phi^2$. \\ \\
The loop integrals contain the functions $A$, $B$ which have divergent
parts that are proportional to a quantity $\Lambda (\mu)$:
\glan \Lambda (\mu)=\frac{\mu^{d-4}}{(4\pi)^2}\gk\frac{1}{d-4}-\frac{1}{2}(\mbox{ln}
(4\pi)+\Gamma'(1)+1)\gK
~,\glaus
\glan A(M^2)=-\frac{M^2}{(4\pi)^2}\,\mbox{ln}\,\frac{M^2}{\mu^2}-2M^2\Lambda
~,\glaus
\glan B(p^2, M^2)=\bar{B}(p^2, M^2)+B(0, M^2) 
\label{bauf}
~,\glaus
\glan B(0, M^2)=-\frac{1}{(4\pi)^2}\,(1+\mbox{ln}\,\frac{M^2}{\mu^2})-2\Lambda
~.\glaus
The quantity $\mu$ is an arbitrary scale with the dimension of mass. With
\glan \hat{\Gamma}^{\mu\nu}=\dmu\gnu-\dnu\gmu+[\gmu, \gnu]
\glaus
the one-loop functional has the following divergence structure \cite{gass1,gass2}:
\glan Z_{\mbox{\scriptsize one loop}}^{\mbox{\scriptsize div}}=
-\frac{\lam}{12}\Tr(\hat{\Gamma}^{\mu\nu}\hat{\Gamma}_{\mu\nu})
-\frac{\lam}{2}\Tr(\hat{\sigma}\hat{\sigma})
\label{ldiv}
~,\glaus
\glan \lam=\Lambda (\mu)+\frac{1}{32\pi^2}\,\mbox{ln}\,\frac{M^2}{\mu^2}
~,\glaus
\glanf \Tr(\hat{\Gamma}^{\mu\nu}\hat{\Gamma}_{\mu\nu})
& =& -\frac{1}{2}\,\kk D_\mu \Ub^\dag D^\mu \Ub \KK^2
+\frac{1}{2}\,\kk D_\mu \Ub^\dag D_\nu \Ub \KK
\kk D^\mu \Ub^\dag D^\nu \Ub \KK+
\nonumber \\
& & -2\,\kk \Ub^\dag F_{\mu\nu}^R \Ub F^{L\,\mu\nu}\KK
-\ii\,\kk F_{\mu\nu}^R\,[D^\mu \Ub, D^\nu \Ub^\dag]+
F_{\mu\nu}^L\,[D^\mu \Ub^\dag, D^\nu \Ub]\KK+
\nonumber \\
& & -\kk F_{\mu\nu}^R F^{\mu\nu R} +F_{\mu\nu}^L F^{\mu\nu L}\KK
\label{ldiv2}
~,\glausf
\glanf  \Tr(\hat{\sigma}\hat{\sigma}) & = & \frac{1}{4}\,\kk D_\mu \Ub^\dag D^\mu \Ub \KK^2
+\frac{1}{4} \,\kk D_\mu \Ub^\dag D_\nu \Ub \KK
\kk D^\mu \Ub^\dag D^\nu \Ub \KK+
\nonumber \\
& & +\frac{1}{2}\,\kk D_\mu \Ub^\dag D^\mu \Ub \KK \kk \chi^\dag \Ub+\chi \Ub^\dag \KK
+\frac{3}{16}\,\kk \chi^\dag \Ub+\chi \Ub^\dag \KK^2
\label{ldiv3}
~,\glausf
\glanf & & \kk D_\mu \Ub^\dag D^\mu \Ub \KK \kk \chi^\dag \Ub+\chi \Ub^\dag \KK=
2\,\kk D_\mu \Ub^\dag D^\mu\chi+ D_\mu \Ub D^\mu\chi^\dag \KK+
\nonumber \\
& & +2\,\kk \chi\chi^\dag+\frac{1}{2}(\chi\tilde{\chi}
-\chi^\dag\tilde{\chi}^\dag)\KK
-\frac{1}{2}\,\kk \chi^\dag \Ub+\chi \Ub^\dag\KK^2
\label{bezieh}
~.\glausf
The divergences can be eliminated by a renormalization of the coupling constants:
\glan l_i= \tilde{l}_i+\gamma_i\,\lam\, , \raum
h_i=\tilde{h}_i+\delta_i\,\lam
~.\glaus
The finite quantities $\tilde{l}_i$ and $\tilde{h}_i$ as well as the divergent $\lam$ are scale independent.
To absorb the divergences of the loops, the $\gamma_i$
and $\delta_i$ take the following values \cite{gass1}:
\glanf & & \gamma_1=\frac{1}{3}\, , \raum  \gamma_2=\frac{2}{3}\, , \raum
\gamma_3=-\frac{1}{2}\, , \raum \gamma_4=2\, , \raum
\gamma_5=-\frac{1}{6}\, , \raum
\nonumber \\
& &\gamma_6=-\frac{1}{3}\, , \raum \gamma_7=0\, , \raum
\delta_1=2\, , \raum \delta_2=\frac{1}{12}\, , \raum
\delta_3=0
~.\glausf
The renormalization scheme amounts to the following replacements:
\glanf  & l_i \pfeil \tilde{l}_i \, , \raum h_i \pfeil \tilde{h}_i \, , & \nonumber \\
& A \pfeil 0 \, , \raum 
B(p^2) \pfeil \bar{B}(p^2)-\frac{1}{(4\pi)^2} &
~.\glausf
All in all we have the renormalized one-loop functional
\glanf
&& \hspace*{-3.5cm}Z_4[U, a, v, s, p]=\int \dx\,\gk {\cal L}_2+{\cal L}^{ren}_4
 \gK+
\nonumber \\
& & \hspace*{-3cm}+\int \dx\,\dy \frac{d^4p}{(2\pi)^4}\,\mbox{e}^{-\iii p (x-y)}\,
\sgk F_1^{\mu\nu}(p) \,
\mbox{Sp}\gk\gmu(x)\gnu(y)\gK+
\nonumber \\
& & \hspace*{-3cm}\raum + F_2(p) \,\mbox{Sp}\gk\sig(x)\sig(y)\gK
+F_3^{\mu\nu}(p)
\,\mbox{Sp}\gk\gmu(x)\gnu(x)\sig(y)\gK\sgK+
\nonumber \\
& & \hspace*{-3cm}+\int \dx\,\dy\,\dz\frac{\dpp}{(2\pi)^4}\frac{\dPP}{(2\pi)^4}
\,\mbox{e}^{\iii p_a (z-x)}\mbox{e}^{\iii p_b (y-z)}
\nonumber \\
& & \hspace*{-3cm}\: \sgk F_4^{\mu\nu\lambda}(p_a,p_b)\,
\mbox{Sp}\gk\gmu(x)\gnu(y)\glambda(z)\gK 
+F_5^{\mu\nu}(p_a,p_b) \:\:\:\mbox{Sp}\gk\gmu(x)\gnu(y)\sig(z)\gK
\nonumber \\
& & \hspace*{-3cm}\raum +F_6^{\mu}(p_a,p_b)
\:\:\:\mbox{Sp}\gk\sig(x)\sig(y)\gmu(z)\gK+F_7 (p_a,p_b)\,\mbox{Sp}\gk\sig(x)\sig(y)\sig(z)\gK\sgK
\label{zdiv}
~.\glausf
\FIGURE{\begin{picture}(100,100)
\put(50,50){\circle{40}}
\put(70,50){\circle*{3}}
\put(30,50){\circle*{3}}
\put(70,50){\line(1,0){20}}
\put(70,50){\vector(1,0){14}}
\put(76.5,42){\makebox(10,10)[b]{\scriptsize $p$}}
\put(30,50){\line(-1,0){20}}
\put(8,50){\vector(1,0){14}}
\put(14,42){\makebox(10,10)[b]{\scriptsize $p$}}
\end{picture}
\begin{picture}(100,100)
\put(50,50){\circle{40}}
\put(35.86,64.14){\circle*{3}}
\put(35.86,35.86){\circle*{3}}
\put(70,50){\circle*{3}}
\put(35.86,64.14){\line(-1,1){14.14}}
\put(35.86,35.86){\line(-1,-1){14.14}}
\put(70,50){\line(1,0){20}}
\put(21.72,78.28){\vector(1,-1){10}}
\put(28.78,71.21){\makebox(8,8)[t]{\scriptsize $p_a$}}
\put(35.86,35.86){\vector(-1,-1){10}}
\put(28.78,18.78){\makebox(8,8)[r]{\scriptsize $p_b$}}
\put(70,50){\vector(1,0){14}}
\put(73,40){\makebox(25,8)[l]{\scriptsize $p_a\! -\!p_b$}}
\end{picture}
\caption{Definition of external momenta.}
\label{momenta}}
${\cal L}^{ren}_4$ is the Lagrangian of  \glei{lop4} with the $l_i$ and $h_i$ replaced by the $\tilde{l}_i$ and
$\tilde{h}_i$.
The scale dependence of the renormalized ${\cal O}(p^4)$-coupling constants has cancelled
the chiral logs of the loops. So everything is scale independent as
it should be.
The constituent functions $F_1^{\mu\nu}\, , \,\dots\, , \, F_7$ can be found in App.~\ref{appa}.
Sp denotes the trace in the space of the 3$\times$3 matrices in the adjoint representation.
The external momenta flow as shown in Fig. \ref{momenta}.

The loop part is of order $p^4$ as one can see by dimensional considerations \cite{wein}.
With ${\cal L}_2$ and ${\cal L}_4$ containing all possible terms respecting the symmetries
of QCD, $Z_4$ is the complete generating functional of ${\cal O}(p^4)$ for Green
functions of quark currents. In the
context of SU(2)$\times$SU(2) there is no anomalous part in the generating functional. \\
\section{Amplitudes for $\gamma\gamma \to \pi\pi$}
\label{ampli1}
In case of pions and photons one can use the following recipe to extract the
scattering amplitudes from the generating functional \cite{gass4}:
\begin{itemize}
\item Expand all quantities which contain the matrices $\bar{U}(\phi)$ or $\bar{u}(\phi)$ in the
pion fields.
\item Replace the external fields $\chi$ by $M$, $a_\mu$ by zero
and $v_\mu$ by $-e\, Q {\cal A}_\mu$  ($Q$ is the quark mass matrix,
${\cal A}_\mu$ is the photon field).
\item Perform functional differentiation with respect to the external pion and photon fields.
\end{itemize}
This method is equivalent to the method described in Sec. \ref{matrix} because the pion matrix
is determined by the equation of motion in the following way
\glan \phi=(\Box+M^2)^{-1}F\, \partial_\mu a^\mu + \dots \, ,
\glaus
and only pole contributions are relevant for scattering amplitudes. \\ \\
The amplitude of the process $\gamma(k_1)\gamma(k_2) \pfeil \pi^0\pi^0$  ($s=(k_1+k_2)^2$) is given by
\glan \M_ {\gamma\gamma \to \pi^0\pi^0}=\frac{4\,e^2}{F_\pi^2}\left(F^{\mu\nu}_3(k_1+k_2)+
F^{\mu\nu}_5(k_1,-k_2)\right)\epsilon_{1\mu}\epsilon_{2\nu}\left(s-M_\pi^2\right)\, .
\glaus
This can be simplified to
\glan \M_ {\gamma\gamma \to \pi^0\pi^0}=
 \epsilon_{1}\cdot\epsilon_{2}
\,\frac{-e^2}{8 \pi^2 F^2_\pi} (s-M^2_\pi)
\left( 1+ \frac{M^2_\pi}{s} \left(\mbox{ln}
\left(\frac{1+\sqrt{1-\frac{4M^2_\pi}{s}}}{1-\sqrt{1-\frac{4M^2_\pi}{s}}}\right)
-\ii\,\pi \right)^2 \right)
\label{pp00}
~.\glaus
The physical quantities $M_\pi, F_\pi$ are obtained through the following
renormalization:
\glanf M_\pi^2&=&M^2(1+\frac{2 M^2}{F^2}\,\tilde{l}_3)+{\cal O}(p^6) \, ,
\nonumber \\
F_\pi&=&F(1+\frac{M^2}{F^2}\,\tilde{l}_4)+{\cal O}(p^6) \, .
\glausf
In this case, where the amplitude starts at order $p^4$, the difference between using $M$, $F$ and
$M_\pi$, $F_\pi$ would appear first at order $p^6$.
The amplitude (\ref{pp00}) is equal to the one in \cite{streuunga}. \\ \\
In case of  $\gamma(k_1)\gamma(k_2) \pfeil \pi^+(p_A)\pi^-(p_B)$ one has:
\glanf & &\M_ {\gamma\gamma \to \pi^+\pi^-}=2\,e^2 \left( a\,\epsilon_1\cdot\epsilon_2
-\frac{p_A\cdot\epsilon_1\,p_B\cdot\epsilon_2}{p_A\cdot k_1}
-\frac{p_A\cdot\epsilon_2\,p_B\cdot\epsilon_1}{p_A\cdot k_2} \right)
~, \glausf
\glanf a&=&1+\frac{s}{F^2_\pi}\,(2\tilde{l}_5-\tilde{l}_6)+\frac{2}{F_\pi^2}\left(F^{\mu\nu}_3(k_1+k_2)+
F^{\mu\nu}_5(k_1,-k_2)\right)\epsilon_{1\mu}\epsilon_{2\nu}\, s \nonumber \\
& = & 1+\frac{s}{F^2_\pi}\,(2\tilde{l}_5-\tilde{l}_6)
-\frac{1}{32\pi^2 F^2_\pi}\left(s+M^2_\pi \left(\mbox{ln}
\left(\frac{1+\sqrt{1-\frac{4M^2_\pi}{s}}}{1-\sqrt{1-\frac{4M^2_\pi}{s}}}\right)
-\ii\,\pi \right)^2 \right) ~.
\nonumber \\
\glausf
This result agrees with the pion part of the amplitude in \cite{streuungb} and
with the ${\cal O}(p^4)$ part
in \cite{streuungc}.
\section{Amplitudes and cross sections for $e^+e^- \to 4\pi$}
\label{ampli2}
The amplitude for the scattering
$$
e^+(k_+) e^-(k_-) \to \pi^0(p_1) \pi^0(p_2) \pi^+(p_3) \pi^-(p_4)
$$
takes the following form
\begin{equation} 
\M_{e^+e^- \to 2\pi^0\pi^+\pi^-} = \displaystyle\frac{e^2}{q^2 + i \ve}\ol v(k_+)\gamma_\mu
u(k_-) J^\mu(p_1,p_2,p_3,p_4) ~,
\hspace*{0.4cm} q=k_+ + k_- = \dsum_{i=1}^4 p_i~,
\end{equation} 
\begin{equation} 
 J^\mu(p_1,p_2,p_3,p_4) := \langle \pi^0(p_1) \pi^0(p_2) \pi^+(p_3) \pi^-(p_4)
|J^\mu_{\rm elm}(0)|0\rangle ~.
\end{equation}
Because of Bose symmetry and $C$ invariance, $J^\mu$ can be written \cite{eu} as
\begin{eqnarray} 
J^\mu(p_1,p_2,p_3,p_4) &=&A^\mu(p_1,p_2,p_3,p_4)+
A^\mu(p_2,p_1,p_3,p_4) \nn
 && - A^\mu(p_1,p_2,p_4,p_3) -
A^\mu(p_2,p_1,p_4,p_3) \label{eq:Adef}
~.
\end{eqnarray}
To next-to-leading order in CHPT, the reduced amplitude $A^\mu(p_1,p_2,p_3,p_4)$ contains the
tree amplitudes of ${\cal O}(p^2)$ and ${\cal O}(p^4)$ and the one-loop part:
\begin{equation}
A^\mu =A^\mu_{(2)}+ A^\mu_{(4){\rm tree}} + A^\mu_{(4){\rm loop}}~.
\end{equation}
In some cases it is convenient to use kinematic variables in the amplitude:
\begin{eqnarray} 
 & s=(p_1 + p_2)^2 ~,  \raum\nu=(p_3 - p_4)\cdot (p_1 - p_2)/2 ~, &\nn
& t_i = p_i\cdot q ~~(i=1,\dots,4)  ~. & 
\end{eqnarray} 
In the following I drop terms proportional to $q^\mu$ because they cannot contribute to the
differential cross section of electroproduction. One can recover them with the help of
current conservation. The amplitude $A^\mu$ is given by:
\begin{equation} 
A^\mu_{(2)}(p_1,p_2,p_3,p_4)= \displaystyle\frac{s-M_\pi^2}
{F_\pi^2} \displaystyle\frac{p_3^\mu}{2 t_3 - q^2} ~,
\label{eq:Ap2}
\end{equation}
\begin{eqnarray} 
F_\pi^4 A^\mu_{(4){\rm tree}}(p_1,p_2,p_3,p_4)& = & 
2 \wt l_2 (\nu-t_3) p_1^\mu \label{eq:tree4} \\
&& \hspace*{-4cm} + \left\{2 \wt l_1 (s^2-4 s M_\pi^2+ 4 M_\pi^4)+
\displaystyle\frac{\wt l_2}{2}(s^2 - 2 t_1 t_2 + 2 t_1^2 - 8 t_1 \nu
+ 4 \nu^2 - (q^2 - 2 t_3)^2) \right. \nn 
&& \hspace*{-4cm}\left. + 2 \wt l_3 M_\pi^4 + 2 \wt l_4 (s M_\pi^2 - M_\pi^4) + 
\wt l_6 q^2 (M_\pi^2 - s) \right\} \displaystyle\frac{p_3^\mu}
{2 t_3 - q^2}~, \nonumber
\end{eqnarray}
\begin{eqnarray} 
F_\pi^4 A^\mu_{(4){\rm loop}}(p_1,p_2,p_3,p_4)& = & 
\frac{2\,\left( 3\,M_\pi^2 + q^2 - 3\,s - 2\,\Mvariable{t_3} \right) \,
    {{{p_3}}_{\nu }}\,{{F_1}}^{\Mvariable{\nu \mu }} (q^2)}
    {3\,\left( 2\,{t_3}-q^2 \right) } \nonumber \\
&& \hspace*{-4.6cm} +\left( \frac{{{{p_3}}^{\mu }}}{2\,{t_3}-q^2} \right)
\sgkg \left[  ({{{p_1}}_{\rho }} - 
     {{{p_3}}_{\rho }}+q_\rho)\,{{{p_2}}_{\nu }}  - 
({{{p_1}}_{\rho }}- 
     {{{p_3}}_{\rho }}+q_\rho)\,{{{p_4}}_{\nu }}  \right] \,
     {{F_1}}^{\Mvariable{\rho \nu }} (q - {p_1} - {p_3})  \nonumber \\
&& \hspace*{-4.6cm}  + 
  \left[ {{{p_1}}_{\rho }}\,({{{p_2}}_{\nu }} - 
     {{{p_3}}_{\nu }}+q_\nu) - 
     {{{p_4}}_{\rho }}\,({{{p_2}}_{\nu }}
     -{{{p_3}}_{\nu }}+q_\nu) \right] \,
     {{F_1}}^{\Mvariable{\rho \nu }} ({p_1} + {p_4})\nonumber \\
&& \hspace*{-4.6cm}  +2\,\gke \left( 3\,M_\pi^4 + 4\,M_\pi^2\,
        \left( -{t_4} + {p_3}\cdot {p_4} \right)  + 
       4\,{p_1}\cdot {p_2}\,\left( M_\pi^2 - {t_4} + 
          {p_3}\cdot {p_4} \right)  \right) \,{F_2}({p_1} + {p_2}) \nonumber \\
&& \hspace*{-4.6cm} +4\,{p_1}\cdot {p_4}\,\left( -{t_2} + {p_2}\cdot {p_3} \right) \,
     {F_2}(-{p_1} - {p_4}) \gKe\sgKg \nonumber \\
&& \hspace*{-4.6cm} -\frac{{{{p_3}}_{\nu }}\,{{F_1}}^{\Mvariable{\mu \nu }} (q)}
   {6} -{{{p_3}}_{\nu }}\, {{F_1}}^{\Mvariable{\mu \nu }} (
    {p_1} + {p_3}) -{{{p_1}}_{\nu }}\,
   {{F_1}}^{\Mvariable{\mu \nu }} ({p_1} + {p_4}) \nonumber \\
&& \hspace*{-4.6cm} -4\,\gke \left( M_\pi^2 + {p_1}\cdot {p_2} \right) \,{{{p_3}}^{\mu }}\,
     {F_2}({p_1} + {p_2}) + {p_1}\cdot {p_3}\,{{{p_2}}^{\mu }}\,
     {F_2}({p_1} + {p_3})  \gKe \nonumber \\
&& \hspace*{-4.6cm} -{p_2}\cdot {p_4}\,\left( {{{p_1}}_{\nu }} - {{{p_3}}_{\nu }} \right)
       \,{{F_3}}^{\Mvariable{\mu \nu }} (q - {p_1} - {p_3}) -
    \left (M_\pi^2+ 2\,{p_1}\cdot {p_2}\right)\,{{{p_3}}_{\nu }}\, {{F_3}}^{\Mvariable{\mu \nu }} (
      q - {p_3} - {p_4})
 \nonumber \\
&& \hspace*{-4.6cm} -\frac{3\,\ii }{4}\,\left( {{{p_1}}_{\nu }}\,{{{p_2}}_{\lambda }} - 
    {{{p_3}}_{\nu }}\,{{{p_2}}_{\lambda }} - 
    {{{p_1}}_{\nu }}\,{{{p_4}}_{\lambda }} + 
    {{{p_3}}_{\nu }}\,{{{p_4}}_{\lambda }} \right) \,
  {{F_4}}^{\Mvariable{\mu \nu \lambda }} (q,{p_1} + {p_3}) \nonumber \\
&& \hspace*{-4.6cm} {- {p_2}\cdot {p_4}\,\left(  {{{p_1}}_{\nu }} - 
       {{{p_3}}_{\nu }} \right) \,
     {{F_5}}^{\Mvariable{\mu \nu }} (q,{p_1} + {p_3}) - 
    \left( M_\pi^2  + 
       2\,{p_1}\cdot {p_2} \right) \,{{{p_3}}_{\nu }}\,
     {{F_5}}^{\Mvariable{\mu \nu }} (q,{p_3} + {p_4})} \nonumber \\
&& \hspace*{-4.6cm}-\ii \,{p_1}\cdot {p_3}\,{p_2}\cdot {p_4}\,
  {{F_6}}^{\mu } (-{p_1} - {p_3},{p_2} + {p_4})\, .
\end{eqnarray} \\
\FIGURE{\epsfig{file=4pxsn4priv.eps,height=6cm}
\caption{Theoretical predictions for the cross section $\sigma(e^+ e^- \to 
2\pi^0\pi^+\pi^-)$ for 0.65 $\le E$(GeV) $\le$ 1.05.}
\label{fig:xsn}}
The replacements $M \to M_\pi$ and $F \to F_\pi$ in the ${\cal O}(p^2)$ part
have given additional contributions to the ${\cal O}(p^4)$ amplitudes.

With the complete amplitude one gets the differential cross section (setting $m_e=0$):
\begin{equation} 
d\sigma = \displaystyle\frac{\alpha^2}{32 \pi^6 q^6}
(\dprod_{i=1}^4\frac{d^3 p_i}{2 E_i})\delta^{(4)}(q-\dsum_{i=1}^4 p_i)
l^{\mu\nu} J_\mu J_\nu^* \label{eq:csec}\, .
\end{equation}
One could perform the same calculations for the process 
$e^+e^-\to 2\pi^+2\pi^-$. In the isospin limit,
which I assume in this article, it is a big simplification to use
instead the following relation \cite{kuehnrev} that
immediately gives the current matrix element for the $2\pi^+2\pi^-$ channel in terms of the matrix element
for the $2\pi^0\pi^+\pi^-$ channel :
\begin{eqnarray} 
\hspace*{-.5cm} \langle \pi^+(p_1) \pi^+(p_2) \pi^-(p_3) \pi^-(p_4)
|J^\mu_{\rm elm}(0)|0\rangle &= & J^\mu(p_1,p_3,p_2,p_4) +
J^\mu(p_1,p_4,p_2,p_3) \nn
& +& J^\mu(p_2,p_3,p_1,p_4) + J^\mu(p_2,p_4,p_1,p_3)~. 
 \label{eq:k2}
\end{eqnarray}
\FIGURE{
\setlength{\unitlength}{1cm}
\begin{picture}(16,8)
\put(0.5,0.03){\makebox(7.0,7.0)[lb]{\epsfig
{figure=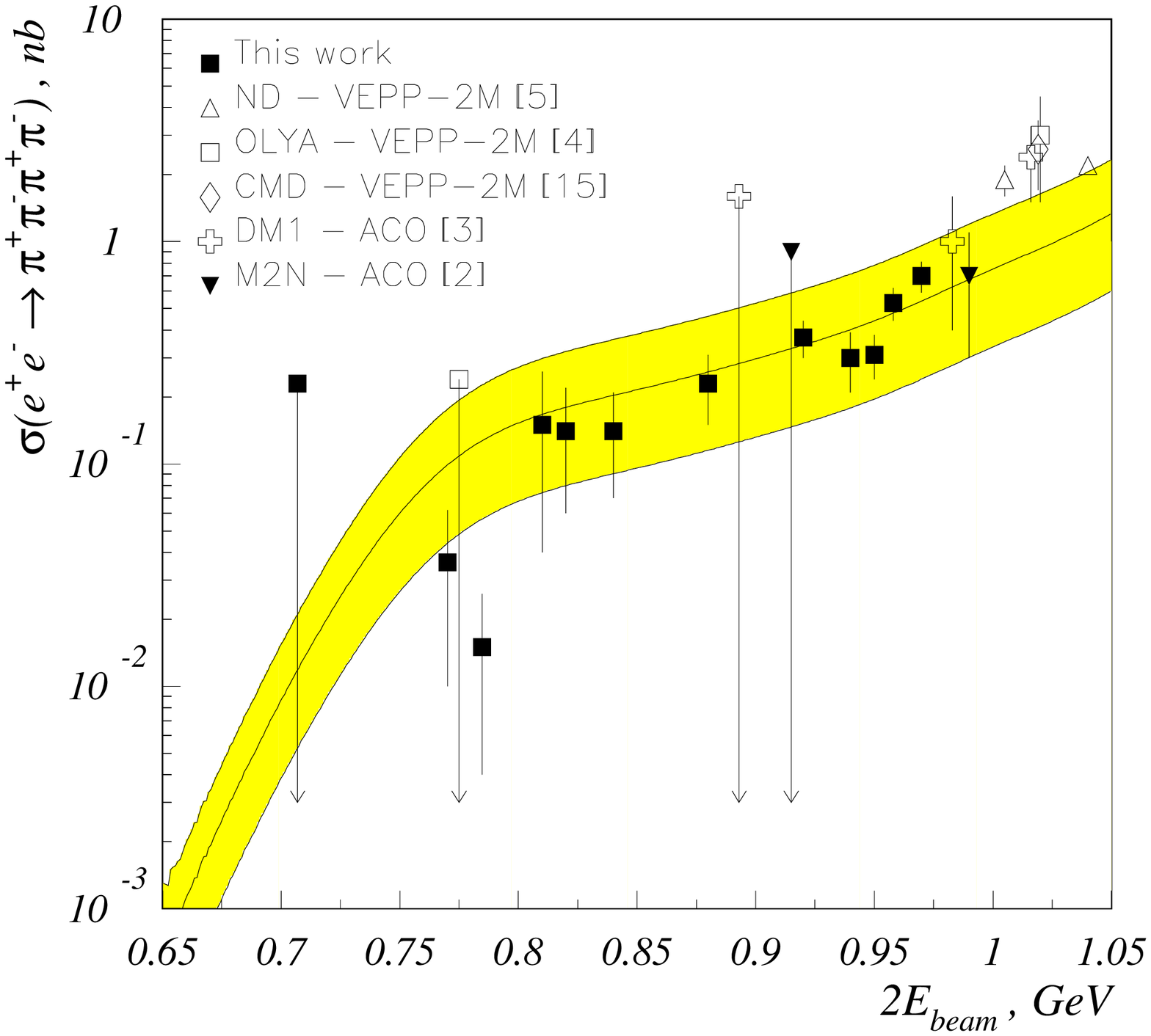,width=6.56cm,height=7.07cm}}}
\put(8.5,0.6){\makebox(7.0,7.0)[lb]
{\epsfig{figure=4pxsc4priv.eps,height=6cm}}}
\end{picture}
\caption{Comparison of data \protect\cite{cmdl} (left figure) and 
predictions (right figure, see text) for the cross section 
$\sigma(e^+ e^- \to 2\pi^+ 2\pi^-)$ for 0.65 $\le E$(GeV) $\le$ 1.05.} 
\label{fig:xsc}}
After numerical integration one gets the cross sections for the two channels. To be able to reach
energies around 1 GeV, resonances have to be included as
described in \cite{eu}. In Fig.~\ref{fig:xsc} and Fig.~\ref{fig:xsn} the dotted curve is the cross 
section for  the $O(p^4)$ amplitude, the full curve corresponds to the full amplitude with
resonance exchange included and the dashed curve represents the complete amplitude without
loops. As one can see, the influence of the loops is very small at higher energies.
A detailed discussion of the results can be found in \cite{eu}.
\section{Conclusions}
\label{conclu}
Chiral perturbation theory is the low-energy effective quantum field theory of QCD.
It is characterized by
chiral symmetry, by the identification of the lightest mesons as Goldstone bosons associated with the
spontaneous breaking of the chiral symmetry and by explicit breaking of chiral symmetry via
the masses of the light quarks.
To parametrize the breaking terms induced by the quark masses, and also to generate in a
systematic way the Green functions of quark currents, it is convenient to insert appropriate
sources in the Lagrangian that promote the global chiral symmetry to a local symmetry.

One has to include loops to satisfy unitarity and to increase the
precision. The loop expansion is an expansion in Planck's constant, i.e 
an expansion around the solution of the classical equation of motion.
The divergences coming from the loops have to be renormalized
by counterterms of the corresponding order in the external momenta.

I gave a compact expression for the generating one-loop functional of chiral SU(2) in the
isospin limit with at most three propagators in the loop. With this functional one can easily get 
all possible amplitudes up to order $\phi^6$, where an external pion is of order $\phi$ and a vector gauge
boson is of order $\phi^2$.

With the help of the generating functional I calculated the processes $e^+ e^-\to \pi^0 \pi^0 \pi^+ \pi^-$
and $e^+ e^-\to 2\pi^+ 2\pi^-$ that are relevant for the hadronic contribution to
the anomalous magnetic moment of the muon and to the fine-structure constant at $M_Z$.
Apart from that I confirmed the existing results for
$\gamma\gamma \pfeil \pi^0\pi^0$ and $\gamma\gamma \pfeil \pi^+\pi^-$
\cite{streuunga,streuungb,streuungc}.
Other processes that can be calculated with the functional would be, for example,
$\gamma^*\pfeil \pi\pi\gamma$ or $\pi\pi \to \pi\pi\pi\pi$.

The corresponding calculation of the generating functional for the
symmetry group SU(3)$\times$SU(3) (which allows also
$K ,\,\eta\,$ to occur as external particles) including the nonleptonic weak interactions \cite{ru2}
is under way. 
\vfill
\section*{Acknowledgements}
\noindent
I thank G. Ecker for supporting me in my work and H. Bijnens for useful informations.
\newpage
\newcounter{zaehler}
\renewcommand{\thesection}{\Alph{zaehler}}
\renewcommand{\theequation}{\Alph{zaehler}.\arabic{equation}}
\setcounter{zaehler}{1}
\setcounter{equation}{0}
\appendix
\section{Constituent functions}
\label{appa}
The constituent functions of the one-loop functional are of the following form:
\glanf & & F_1^{\mu\nu}(p)=p^\mu p^\nu \, a(p^2)+ g^{\mu\nu} \, b(p^2)\, ,
\nonumber \\ \nonumber \\
& & F_2(p)=\frac{1}{4} B(p^2) \, ,\raum
F_3^{\mu\nu}(p)= \frac{1}{2} g^{\mu\nu} B(p^2)\, ,\nonumber \\ \nonumber \\
& & F_4^{\mu\nu\lambda}(p_a,p_b)=\ii\,\gk c\aaa\ppm\ppn\ppl +c\bbb\PPm\PPn\PPl
+d\aaa\ppm\ppn\PPl
\nonumber \\
& & \raum +e\aaa\ppm\PPn\ppl+f\aaa\PPm\ppn\ppl
+d\bbb\PPm\PPn\ppl+f\bbb\PPm\ppn\PPl 
\nonumber \\
& & \raum  +e\bbb\ppm\PPn\PPl
+g\aaa\ppm g^{\nu\lambda}+h\aaa\ppn g^{\mu\lambda}+i\aaa\ppl g^{\mu\nu} 
\nonumber \\
& & \raum +h\bbb\PPm g^{\nu\lambda}+g\bbb\PPn g^{\mu\lambda} +i\bbb\PPl g^{\mu\nu}\gK \, ,
\nonumber \\ \nonumber \\
& & F_5^{\mu\nu}(p_a,p_b)=j\aaa g^{\mu\nu} +k\aaa\ppm\ppn+k\bbb\PPm\PPn
\nonumber \\
& & \raum  +l\aaa\ppm\PPn+m\aaa\PPm\ppn \, ,
\nonumber \\ \nonumber \\
& & F_6^{\mu}(p_a,p_b)=\ii\,\gk n\aaa\ppm +n\bbb\PPm \gK\, , \raum F_7(p_a,p_b)=\frac{1}{6}\,C\, ,
\glausf
with for example
\glan c_a=c(p_a^2, p_b^2, p_a\cdot p_b)\raum\mbox{and}
\raum c_b=c(p_b^2, p_a^2, p_a\cdot p_b) \, ,
\glaus
where
\glanf
&&a(p^2)=
\frac{{B}(p^2)-4\,{B_{22}}(p^2)}
  {4}  \, , \raum
b(p^2)=
-{B_{20}}(p^2) \, ,\nonumber \\ \nonumber \\
&&c_a=
\frac{{C_{11\,a}}  - 
    4\,C_{22\,a}+ 4\,{C_{33\,a}}}{3} \, , \raum
d_a=
\frac{ {C_{11\,a}} - 
    2\,C_{22\,a} -2\,{{\tilde{C}}_{22}}  + 
    4\,{{\tilde{C}}}_{33\,a}}{3} \, ,\nonumber \\ \nonumber \\
&& e_a=
\frac{ -C/2+2\,{C_{11\,a}} + 
    {C_{11\,b}} - 
    2\,C_{22\,a}-4\,{{\tilde{C}}_{22}} + 
    4\,{{\tilde{C}}}_{33\,a} }{3}  \, ,\nonumber \\ \nonumber \\
&& f_a=
\frac{-2\,\left( {{\tilde{C}}_{22}} - 
      2\,{{\tilde{C}}}_{33\,a} \right) }{3}  \, , \raum
g_a=
\frac{-2\,\left( {C_{20}} - 2\,{C_{31\,a}} \right) }{3} \, , \raum
h_a=
\frac{4\,{C_{31\,a}}}{3} \, ,\nonumber \\ \nonumber \\
&& i_a=
\frac{-2\,\left( {C_{20}} - 2\,{C_{31\,a}} \right) }{3} \, , \raum
j_a=
-2\,{C_{20}} \, , \raum
k_a=
{C_{11\,a}} - 2\,C_{22\,a} \, , \nonumber \\ \nonumber \\
&& l_a=
- \frac{{C}}{2} + {C_{11\,a}} + {C_{11\,b}}-2\,{{\tilde{C}}_{22}} \, , \raum
m_a=
-2\,{{\tilde{C}}_{22}} \, , \raum
n_a=
\frac{ {C}}{2}-2\,{C_{11\,a}} 
~.\glausf
The B and C functions can be found in App.~\ref{oli}.
\section{One-loop integrals}
\label{oli}
The functions $B$, $B_{ij}$, $C$ and $C_{ij}$ are defined through the following
relations \cite{hooft, pass, kniehl} ($C_F$ is the Feynman integration contour):
\glan \{X\} \doteq \frac{1}{\ii}\int_{C_F}\ddk\frac{X}{(k^2-M^2)({(k-p)}^2-M^2)}
~,\glaus
\glan B(p^2)=\{ 1\}
~,\glaus
\glan \{k_\mu k_\nu\}=g_{\mu\nu}B_{20}(p^2)+p_\mu p_\nu B_{22}(p^2)
~,\glaus
\glan \{\{X\}\} \doteq \frac{1}{\ii}\int_{C_F}\ddk
\frac{X}{(k^2-M^2)({(k-p_a)}^2-M^2)({(k-p_b)}^2-M^2)}
~,\glaus
\glan \{\{k_\mu\}\}=p_{a\mu}\,C_{11a}+p_{b\mu}\,C_{11b}
~,\glaus
\glan C=\{\{1\}\}~,
\glaus
\glan
\{\{k_\mu k_\nu\}\}=g_{\mu\nu}\,C_{20}+p_{a\mu}p_{a\nu}\,C_{22a}+
p_{b\mu}p_{b\nu}\,C_{22b}+(p_{a\mu}p_{b\nu}+p_{b\mu}p_{a\nu})\,\tilde{C}_{22}
~,\glaus
\glanf \{\{k_\mu k_\nu k_\lambda \}\} &=& p_{a\mu}p_{a\nu}p_{a\lambda}\,C_{33a}+
p_{b\mu}p_{b\nu}p_{b\lambda}\,C_{33b}+
\nonumber \\
& & +(p_{a\mu}p_{a\nu}p_{b\lambda}+p_{a\mu}p_{b\nu}p_{a\lambda}
+p_{b\mu}p_{a\nu}p_{a\lambda})\,\tilde{C}_{33a}+
\nonumber \\
& & +(p_{b\mu}p_{b\nu}p_{a\lambda}+p_{b\mu}p_{a\nu}p_{b\lambda}
+p_{a\mu}p_{b\nu}p_{b\lambda})\,\tilde{C}_{33b}+
\nonumber \\
& & +(p_{a\mu}g_{\nu\lambda}+p_{a\nu}g_{\mu\lambda}+p_{a\lambda}g_{\mu\nu})\,C_{31a}+
\nonumber \\
& & +(p_{b\mu}g_{\nu\lambda}+p_{b\nu}g_{\mu\lambda}+p_{b\lambda}g_{\mu\nu})\,C_{31b}
~,\glausf  \\
with $C_{ij\,a}=C_{ij}(p_a^2, p_b^2, p_a\cdot p_b)$, $C_{ij\,b}=C_{ij}(p_b^2, p_a^2, p_a\cdot p_b)$
and $C_{ij}=C_{ij\,a}=C_{ij\,b}$. The explicit form of the functions is
\glan A(M^2)=M^2\left( B(0,M^2)+\frac{1}{16\pi^2}\right) \, , \glaus
$$ B(p^2,M^2)=\bar{B}(p^2,M^2)+B(0,M^2) \, ,  $$
\glan  \bar{B}(p^2,M^2)=\frac{1}{16\pi^2}\,\left(\sqrt{1-\frac{4(\MM)}{p^2}}\,\mbox{ln}\,
\frac{\sqrt{1-\frac{4(\MMM)}{p^2}}-1}{\sqrt{1-\frac{4(\MMM)}{p^2}}+1}+2\right) \, ,
\glaus \\
$$ C(p_a^2, p_b^2, p_a\cdot p_b,M^2)=-\frac{1}{16\pi^2\sqrt{\lambda}}\sum^3_{i=1}\sum_{\sigma=\pm}
\left(\mbox{Li}_2\left(\frac{x_i}{x_i-z_i^\sigma}\right)-
\mbox{Li}_2\left(\frac{x_i-1}{x_i-z_i^\sigma}\right)\right)\, ,   $$
$$ \lambda=4\left( (p_a\cdot p_b)^2-p_a^2 p_b^2\right) \, , \raum
 p_D^2=p_a^2-2p_a\cdot p_b+p_b^2 \, ,  $$
\glan x_1=\frac{1}{2}\left(1+\frac{p_a^2-p_b^2-p_D^2}{\sqrt{\lambda}}\right) \, , \raum
z_1^\sigma=\frac{1}{2}\left(1\pm\sqrt{1-\frac{4}{p_a^2}(M^2-\ii\,\epsilon)}\right) \, .
\glaus
The quantities $x_2$, $z_2^\sigma$ are obtained by the interchange
$p_a^2 \leftrightarrow p_D^2$ and $x_3$, $z_3^\sigma$ by the interchange
$p_a^2 \leftrightarrow p_b^2$. Li$_2$ denotes the dilogarithm
\glan \mbox{Li}_2(z)=-\int_0^z \frac{dt}{t}\mbox{ln}\,(1-t) \, .
\glaus \\
The coefficients functions are of the form
$$
{B_{20}}(p^2) = \frac{6\,M^2 - p^2 + 48\,{\pi }^2\,A + 
     24\,{\pi }^2\,\left( 4\,M^2 - p^2 \right) \,B(p^2)}{288\,{\pi }^2}
\, , $$
\glan
{B_{22}}(p^2) = \frac{-6\,M^2 + p^2 + 96\,{\pi }^2\,A - 
     96\,{\pi }^2\,\left( M^2 - p^2 \right) \,B(p^2)}{288\,p^2\,{\pi }^2}
\, . \glaus \\ \\
In the following $B(1)$ is defined as $B(p_b^2)$ and $B(2)$ as $B(p_a^2-2\,\qqo+p_b^2)$. The same is valid
for $B_{20}$ and $B_{22}$.  \\
$$ {C_{11\,a}} = \frac{ 
     {H_{11\,a}}\,{{p_b}}^2- {H_{11\,b}}\,\Mvariable{\qqo}}{{{p_a}}^2\,{{p_b}}^2-{\Mvariable{\qq}}^2}
\, , $$ $$
{C_{20}} = \frac{2\,C\,M^2 + \frac{1}{16\,{\pi }^2} + B(2) - 
     {C_{11\,a}}\,{{p_a}}^2 - {C_{11\,b}}\,{{p_b}}^2}{4}
\, , $$ $$
{C_{22\,a}} = \frac{ 
     {H_{21\,a}}\,{{p_b}}^2- {H_{22\,b}}\,\Mvariable{\qqo}}{{{p_a}}^2\,{{p_b}}^2-{\Mvariable{\qq}}^2}
\, , \raum {{\tilde{C}}_{22}} = \frac{ 
     {H_{22\,b}}\,{{p_a}}^2- {H_{21\,a}}\, \Mvariable{\qqo}}{{{p_a}}^2\,{{p_b}}^2-{\Mvariable{\qq}}^2}
\, , $$ $$
{C_{33\,a}} = \frac{ 
     {H_{31\,a}}\,{{p_b}}^2-{H_{32\,b}} \,\Mvariable{\qqo}}{{{p_a}}^2\,{{p_b}}^2-{\Mvariable{\qq}}^2 }
\, , \raum {{{\tilde{C}}}_{33\,a}} = \frac{ 
     {H_{32\,b}}\,{{p_a}}^2- {H_{31\,a}} \,\Mvariable{\qqo}}{{{p_a}}^2\,{{p_b}}^2-{\Mvariable{\qq}}^2}
\, , $$ $$
{C_{31\,a}} = \frac{ 
     {H_{30\,a}}\,{{p_b}}^2-{H_{30\,b}} \,\Mvariable{\qqo}}{{{p_a}}^2\,{{p_b}}^2-{\Mvariable{\qq}}^2}
\, , $$ \\  $$
{H_{11\,a}} = \frac{-B(1) + B(2) + C\,{{p_a}}^2}{2}
\, , \raum {H_{21\,a}} = -{C_{20}} + \frac{{B(2)}/{2} + {C_{11\,a}}\,{{p_a}}^2}{2}
\, , $$ $$
{H_{22\,a}} =\frac{ [{-B(1) + B(2)}]/{2} + {C_{11\,b}}\,{{p_a}}^2}{2}
\, , \raum {H_{30\,a}} = \frac{{C_{20}}\,{{p_a}}^2 - {B_{20}}(1) + {B_{20}}(2)}{2}
\, , $$ $$
{H_{31\,a}} = -2\,{C_{31\,a}} + \frac{{{p_a}}^2\,{C_{22\,a}} + 
      {B_{22}}(2)}{2}
\, , \raum {H_{32\,a}} = \frac{{{p_a}}^2\,{C_{22\,b}} - {B_{22}}(1) + 
     {B_{22}}(2)}{2}
\, . $$
\glan \glaus
The divergent functions need to be renormalized as described in
section \ref{onefunk}.
\end{document}